\documentclass[doublespacing,longauth]{elsart}
\usepackage{ifpdf}
\usepackage{graphicx,natbib,amssymb,lineno}
\usepackage{amsmath}  
\usepackage{wasysym}  
\usepackage{epsfig}
\usepackage{textcomp}

\makeatletter
\def\elsartstyle{%
    \def\normalsize{\@setfontsize\normalsize\@xiipt{14.5}}
    \def\small{\@setfontsize\small\@xipt{13.6}}
    \let\footnotesize=\small
    \def\large{\@setfontsize\large\@xivpt{18}}
    \def\Large{\@setfontsize\Large\@xviipt{22}}
    \skip\@mpfootins = 18\p@ \@plus 2\p@
    \normalsize
}
\@ifundefined{square}{}{\let\Box\square}
\makeatother

\begin{document}

\begin{frontmatter}
\begin{center}
\title{Very high energy gamma-ray observations during moonlight and
twilight with the MAGIC telescope} 
\end{center}

\author[a]{J.~Albert},
\author[b]{E.~Aliu},
\author[c]{H.~Anderhub},
\author[d]{P.~Antoranz},
\author[b]{A.~Armada},
\author[e]{C.~Baixeras},
\author[d]{J.A.~Barrio},
\author[g]{H.~Bartko},
\author[h]{D.~Bastieri},
\author[f]{J.~K.~Becker},
\author[j]{W.~Bednarek},
\author[a]{K.~Berger},
\author[h]{C.~Bigongiari},
\author[c]{A.~Biland},
\author[g,h]{R.K.~Bock},
\author[s]{P.~Bordas},
\author[s]{V.~Bosch-Ramon},
\author[a]{T.~Bretz},
\author[c]{I.~Britvitch},
\author[d]{M.~Camara},
\author[g]{E.~Carmona},
\author[k]{A.~Chilingarian},
\author[l]{S.~Ciprini},
\author[g]{J.~A.~Coarasa},
\author[c]{S.~Commichau},
\author[d]{J.L.~Contreras},
\author[b]{J.~Cortina},
\author[u]{M.T.~Costado},
\author[f]{V.~Curtef},
\author[k]{V.~Danielyan},
\author[h]{F.~Dazzi},
\author[u]{C.~Delgado},
\author[i]{A.~De~Angelis},
\author[d]{R.~de~los~Reyes},
\author[i]{B.De~Lotto},
\author[b]{E.~Domingo-Santamar\'\i a},
\author[a]{D.~Dorner},
\author[h]{M.~Doro},
\author[b]{M.~Errando},
\author[o]{M.~Fagiolini},
\author[n]{D.~Ferenc},
\author[b]{E.~Fern\'{a}ndez},
\author[b]{R.~Firpo},
\author[b]{J.~Flix},
\author[d]{M.V.~Fonseca},
\author[e]{L.~Font},
\author[g]{M.~Fuchs},
\author[g]{N.~Galante},
\author[u]{R.~Garc\'{\i}a-L\'opez},
\author[g]{M.~Garczarczyk},
\author[h]{M.~Gaug},
\author[j]{M.~Giller},
\author[g]{F.~Goebel},
\author[k]{D.~Hakobyan},
\author[g]{M.~Hayashida},
\author[m]{T.~Hengstebeck},
\author[u]{A.~Herrero},
\author[a]{D.~H\"ohne},
\author[g]{J.~Hose},
\author[g]{C.~C.~Hsu},
\author[j]{P.~Jacon},
\author[g]{T.~Jogler},
\author[m]{O.~Kalekin},
\author[g]{R.~Kosyra},
\author[m]{D.~Kranich},
\author[a]{R.~Kritzer},
\author[n]{A.~Laille},
\author[g]{P.~Liebing},
\author[l]{E.~Lindfors},
\author[h]{S.~Lombardi},
\author[p]{F.~Longo},
\author[b]{J.~L\'{o}pez},
\author[d]{M.~L\'{o}pez},
\author[c,g]{E.~Lorenz},
\author[g]{P.~Majumdar},
\author[p]{G.~Maneva},
\author[a]{K.~Mannheim},
\author[i]{O.~Mansutti},
\author[h]{M.~Mariotti},
\author[b]{M.~Mart\'{i}nez},
\author[g]{D.~Mazin},
\author[g]{C.~Merck},
\author[o]{M.~Meucci},
\author[a]{M.~Meyer},
\author[d]{J.~M.~Miranda},
\author[g]{R.~Mirzoyan},
\author[g]{S.~Mizobuchi},
\author[b]{A.~Moralejo},
\author[l]{K.~Nilsson},
\author[g]{J.~Ninkovic},
\author[b]{E.~O\~{n}a-Wilhelmi\corauthref{cor1}},
\corauth[cor1]{Corresponding author.}
\ead{emma@ifae.es}
\author[g]{N.~Otte},
\author[d]{I.~Oya},
\author[g]{D.~Paneque},
\author[u]{M.~Panniello},
 \author[o]{R.~Paoletti},
\author[s]{J.~M.~Paredes},
\author[l]{M.~Pasanen},
\author[h]{D.~Pascoli},
\author[c]{F.~Pauss},
\author[o]{R.~Pegna},
\author[i,q]{M.~Persic},
\author[h]{L.~Peruzzo},
\author[o]{A.~Piccioli},
\author[a]{M.~Poller},
\author[b]{N.~Puchades},
\author[h]{E.~Prandini},
\author[k]{A.~Raymers},
\author[f]{W.~Rhode},
\author[s]{M.~Rib\'o},
 \author[b]{J.~Rico\corauthref{cor2}},
\corauth[cor2]{Corresponding author.}
\ead{jrico@ifae.es}
\author[c]{M.~Rissi},
\author[e]{A.~Robert},
\author[a]{S.~R\"ugamer},
\author[h]{A.~Saggion},
\author[e]{A.~S\'{a}nchez},
 \author[h]{P.~Sartori},
 \author[h]{V.~Scalzotto},
\author[i]{V.~Scapin},
\author[a]{R.~Schmitt},
\author[g]{T.~Schweizer},
\author[m,g]{M.~Shayduk},
\author[g]{K.~Shinozaki},
\author[r]{S.~N.~Shore},
 \author[b]{N.~Sidro},
 \author[l]{A.~Sillanp\"{a}\"{a}},
 \author[j]{D.~Sobczynska},
 \author[o]{A.~Stamerra},
 \author[c]{L.~Stark},
 \author[l]{L.~Takalo},
 \author[p]{P.~Temnikov},
\author[b]{D.~Tescaro},
\author[g]{M.~Teshima},
\author[g]{N.~Tonello},
\author[b,t]{D.~F.~Torres},
\author[o]{N.~Turini},
\author[p]{H.~Vankov},
\author[i]{V.~Vitale},
\author[g]{R.M.~Wagner},
\author[j]{T.~Wibig},
\author[g]{W.~Wittek},
\author[h]{F.~Zandane},
\author[b]{R.~Zanin},
\author[e]{J.~Zapatero}

\address[a] {Universit\"at W\"urzburg, D-97074 W\"urzburg, Germany}
\address[b] {Institut de F\'\i sica d'Altes Energies, Edifici Cn., E-08193 Bellaterra (Barcelona), Spain}
\address[c] {ETH Zurich, CH-8093 Switzerland}
\address[d] {Universidad Complutense, E-28040 Madrid, Spain}
\address[e] {Universitat Aut\`onoma de Barcelona, E-08193 Bellaterra, Spain}
\address[f] {Universit\"at Dortmund, D-44227 Dortmund, Germany}
\address[g] {Max-Planck-Institut f\"ur Physik, D-80805 M\"unchen, Germany}
 \address[h] {Universit\`a di Padova and INFN, I-35131 Padova, Italy} 
 \address[i] {Universit\`a di Udine, and INFN Trieste, I-33100 Udine, Italy} 
 \address[j] {University of \L \'od\'z, PL-90236 Lodz, Poland} 
 \address[k] {Yerevan Physics Institute, AM-375036 Yerevan, Armenia}
 \address[l] {Tuorla Observatory, Turku University, FI-21500 Piikki\"o, Finland}
 \address[m] {Humboldt-Universit\"at zu Berlin, D-12489 Berlin, Germany} 
 \address[n] {University of California, Davis, CA-95616-8677, USA}
 \address[o] {Universit\`a  di Siena, and INFN Pisa, I-53100 Siena, Italy}
\address[p]  {Institute for Nuclear Research and Nuclear Energy, BG-1784 Sofia, Bulgaria}
 \address[q] {INAF/Osservatorio Astronomico and INFN Trieste, I-34131 Trieste, Italy} 
 \address[r] {Universit\`a  di Pisa, and INFN Pisa, I-56126 Pisa, Italy}
 \address[s] {Universitat de Barcelona, E-08028 Barcelona, Spain}
\address[t]  {Institut de Ciencies de l'Espai (IEEC-CSIC), Torre C5 Parell, E-08193 Bellaterra (Barcelona), Spain}
\address[u]  {Instituto de Astrofisica de Canarias, E-38200, La Laguna, Tenerife, Spain}

\begin{abstract} %

We study the capability of the MAGIC telescope to observe under
moderate moonlight. TeV $\gamma$-ray signals from the Crab nebula were
detected with the MAGIC telescope during periods when the Moon was
above the horizon and during twilight. This was accomplished by
increasing the trigger discriminator thresholds. No change is
necessary in the high voltage settings since the camera PMTs were
especially designed to avoid high currents. We characterize the
telescope performance by studying the effect of the moonlight on the
$\gamma$-ray detection efficiency and sensitivity, as well as on the
energy threshold.

\end{abstract}

\end{frontmatter}

\section{Introduction} %

Ground-based searches for very high energy (VHE) $\gamma$-ray emission
from celestial objects are normally carried out by so-called imaging
air Cherenkov telescopes (IACT) during clear, moonless nights. The MAGIC
IACT~\cite{Lorenz} has been designed to carry out observations also
during moderate moonlight. In this paper we describe the technical
innovations and analysis changes that allow observations in the
presence of the Moon.

The MAGIC (Major Atmospheric Gamma Imaging Cherenkov) Telescope is
located on the Canary Island La Palma (2200~m asl, $28^\circ45' N$,
$17^\circ54'W$). MAGIC is currently the largest IACT, with a 17
m-diameter tessellated reflector dish~\cite{Cortina}. The camera is
equipped with 576 6-dynode compact photo-multiplier tubes\footnote{\it
type 9116 $\O$ 25.4~mm, and 9117 of $\O$ 34~mm, from Electron Tubes
Inc. with CsRb cathodes with a peak QE of around 26$\%$ and spectral
sensitivity extended to 650~nm.}(PMTs) with enhanced quantum
efficiency~\cite{Paneque}. The total field of view is 3.5$^\circ$,
divided into two sections: an inner hexagon of 396 small pixels, of
about 3~cm (0.1$^\circ$) diameter, which also corresponds roughly to
the trigger area, and the outer rings of 6~cm diameter pixels. The use
of larger pixels in the outer zone reduces the cost of the camera,
while the quality of the image, already limited in this zone by coma
aberration, it not deteriorated.

One of the unique features of MAGIC is its capability to observe under
moderate moonlight. MAGIC has an average duty cycle per year of about
12$\%$ under strict condition of dark observations, i.e.\ between
astronomical dusk and dawn and with the Moon below the horizon. If
this strict requirement is relaxed to allow observations under
moderate moonlight or twilight, an increase of the duty cycle to
18$\%$ (from $\sim$1000 to $\sim$1500 hours of observations per year)
is possible. Such an increase in the duty cycle places MAGIC in a
prominent position, in particular for the study of variable sources as
well as in multi-wavelength campaigns together with other instruments.

Observations during moonlight were tested by the Whipple collaboration
\cite{Pare}. Their approach was based on restricting the PMTs
sensitivity to the UV range, either by using solar blind PMTs or by
using UV sensitive filters in front of regular PMTs
\cite{Chantell}. Later, the CT1 telescope of the predecessor experiment HEGRA
pioneered regular operations under moderate moonlight
\cite{Kranich}, by lowering the PMT high voltages (HV). Regular
observations were done with CT1 under such conditions, during its last
years of operation. All these solutions were feasible, but also
expensive, time consuming and not efficient in terms of energy
threshold and sensitivity. Also, the change of the PMTs or of the
filters were cumbersome during nights partially dark and partially
with moonlight.

In section~\ref{sec:pmt} we present the technical aspects regarding
observations under moonlight with MAGIC. Section~\ref{sec:data}
describes the data used in this study, the analysis technique and some
technical aspects related to the increase of the
moonlight. Section~\ref{sec:hillas} discusses the effects of the
moonlight on the reconstruction of the shower images, based on Crab
nebula data. In section~\ref{sec:results} we present the telescope
response in terms of $\gamma$-ray detection efficiency, sensitivity
and energy threshold. The conclusions and some recommendations for
moonlight data acquisition are given in Section~\ref{sec:concl}.

\section{Technical consideration for MAGIC observations under the moonlight} %
\label{sec:pmt}

\begin{figure}[!p]
\begin{center}
\epsfig{file=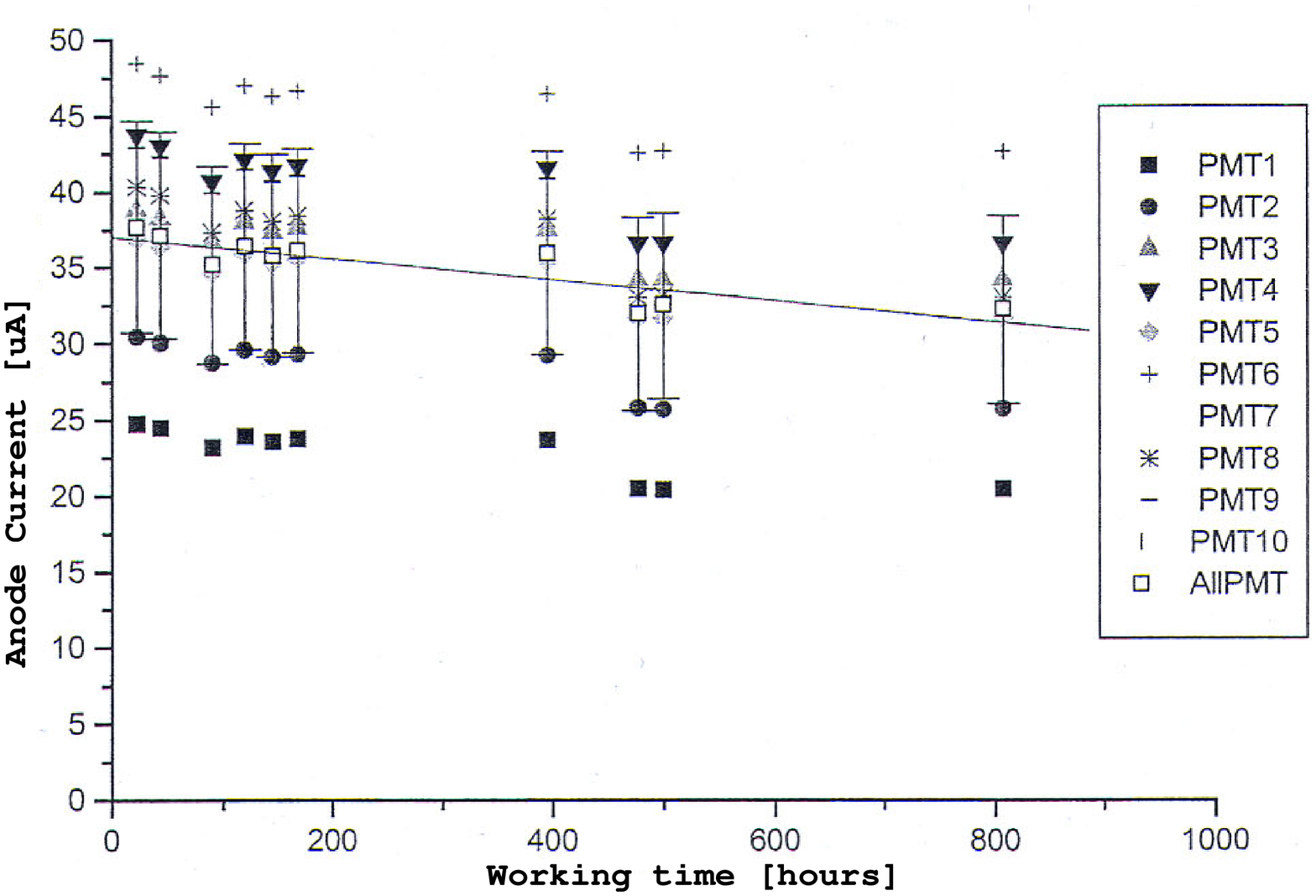,width=0.8\textwidth}
\caption{{\it Ageing studies for a set of ten MAGIC PMTs. The figure shows
the drop in the anode current as a function of time. The data and error bars for
``ALL PMTs'' are the mean value and RMS, respectively, for all PMTs.}}
\label{ageing}
\end{center}
\end{figure}

Traditionally, PMTs are operated with amplification gains around
$10^6-10^7$ which, under moonlight, generate continuous (direct)
currents (DCs) that can damage the last dynode, resulting in rapid
ageing of the PMT. In addition very high anode currents liberate many
absorbed molecules (mostly water) and ionize them. Quite a few of the
ions might diffuse into the PMT front-end volume and get accelerated
towards the cathode where they liberate a large number of electrons on
impact, thus generating large secondary pulses (so-called
afterpulses). The PMT ageing is dominantly caused by a damage of the
last dynode due to this intense electron bombardment, which strongly
reduces the electron emission of these dynodes. The reduction depends
normally on the total charge per unit dynode area, as well as on the
dynode material and its production characteristics. Due to the
electron multiplication in the dynodes, obviously the last dynode is
the most affected. The gain of a 1~cm$^2$ area CuBe dynode drops by a
factor two for an integrated charge of 200-400 Coulomb. In most cases
the gain drop due to the ageing of the last dynode can easily be
compensated by increasing the HV thus resetting the gain of the PMT
back to the initial value. In order to study the range of ageing of
the MAGIC PMTs under the influence of scattered moonlight an
accelerated test has been carried out for a sample of ten of the
Electron Tubes 9116 B PMTs which are used in MAGIC~\cite{PMTs}. The
PMTs, operated at the standard MAGIC electronic conditions, were
exposed to light from a stabilized red LED resulting initially in a
mean anode current of 37~$\mu$A. The anode current drop was monitored
regularly over a time period of 820~h. Figure~\ref{ageing} shows the
gain drop, averaged over the 10 PMTs, for the duration of the test. We
observed a gain drop of $\approx$ 19$\%$ per 1000~h operation at this
high level of illumination. The integrated anode charge was 120
Coulomb. The maximum moonlight illumination which is allowed in MAGIC
corresponds to an anode current of 8~$\mu$A. For this anode current we
estimate a gain drop of 3.8$\%$ per 1000~h observation time. This time
is more than a typical accumulated observation time per year
during bright moonlight (here we make the assumption that no HV
reset to recover the gain occurs during the gaps between nightly and
monthly observations). For compensating this ageing effect we
regularly readjust the HV of the MAGIC PMTS once or twice a year.
From the accelerated ageing study and practical experience over more
than two years one can conclude that the moonlight operation of low
gain, six dynode PMTs is very safe and requires rarely modest gain
adjustments. Actually, it is possible to operate the PMTs without
serious degradation in the presence of much brighter moonlight.

The MAGIC camera was designed to allow observations under different
light conditions, with no need of lowering the HVs. It is the only
Cherenkov telescope camera equipped with PMTs that run at a gain of
about 3$\times10^4$~\cite{Performance} thus avoiding high anode
currents. In order to also detect single photoelectrons (phe) the PMT
signal is fed to an AC-coupled fast, low noise preamplifier to raise
the combined gain to about $10^6$. The DC anode currents are in first
order proportional to the photoelectron rate and in turn to the photon
rate. When observing at a dark area of the night sky the anode
currents are typically 0.8~$\mu$A. This corresponds to a night sky
background light of $1.7\times 10^{12}$~ph~m$^{-2}$~s$^{-1}$~sr$^{-1}$
\cite{Mirzoyan}. Direct moonlight during full Moon is about a few times
$10^{15}$~ph~m$^{-2}$~s$^{-1}$. The increase of the background light
due to the presence of the Moon depends on various factors including
the source zenith angle, Moon phase, angular distance to the Moon,
Moon zenith angle, and atmospheric composition as well as the aerosol
content of the atmosphere. Even a clear atmosphere results in a loss
of 10-20$\%$ of the light, mostly by Rayleigh scattering. Unavoidable
Mie scattering results in strong light intensity close to the Moon
direction. At around 25$^\circ$ away from the Moon the direct
scattered moonlight approaches a constant level below the level of the
night sky light background. In addition, moonlight has a spectral
distribution different from that of the light of the night sky (LONS),
peaking at blue wavelengths and thus better matching the Cherenkov
light than the dark night LONS.

We restrict MAGIC observations to a maximum DC of 8~$\mu$A. This permits
observations in the presence of the Moon until (since) 3-4 days before
(after) full Moon, for an angular distance to the Moon greater than
50$^\circ$~\cite{Performance}.

Like most other IACTs, MAGIC operates on a double trigger threshold
\cite{trigger}, i.e., the hardware trigger threshold which is
determined by the fluctuations in the LONS, and a higher software
threshold for image reconstruction. A typical trigger condition
requires that a minimum number of neighboring pixels (e.g. four in the
present MAGIC configuration) exhibit a signal larger than a given
threshold of a few phe within a short time window.

\begin{figure}[!t]
\begin{center}
\epsfig{file=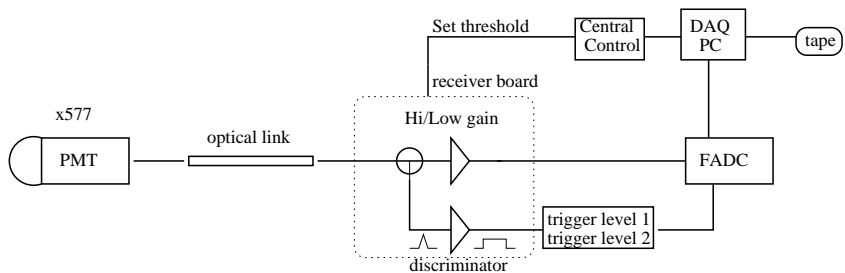,width=0.8\textwidth}
\caption{{\it Sketch of the MAGIC signal propagation chain.}}
\label{sketch}
\end{center}
\end{figure}

The PMT analog signal (see Figure~\ref{sketch}) is
transmitted over an optical fiber, converted into an electrical pulse
and split into two branches. One of the branches is routed to the
digitizers (FADCs). A second branch enters a discriminator, which
issues a digital signal (5.5~ns FWHM) whenever the pulse exceeds a
given threshold. The discriminator thresholds (DT) are set by an 8-bit
digital-to analog converter (DAC) which is controlled from a PC. The
thresholds can be modified during telescope operation.

\begin{figure}[t]
\begin{center}
\epsfig{file=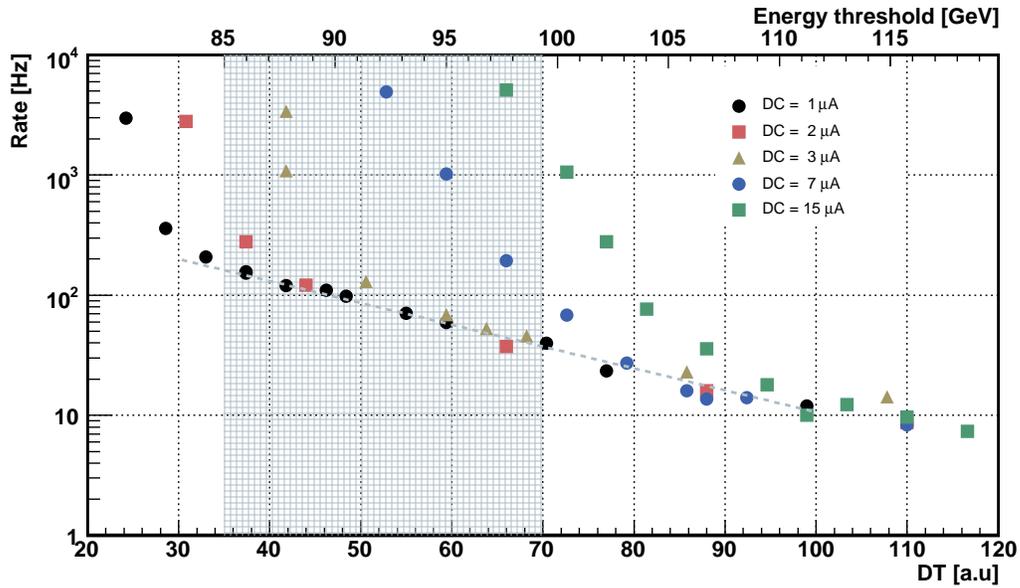,width=\textwidth}
\caption{\it Trigger rate as a function of the discriminator
threshold for four neighboring pixels configuration and different
camera illuminations. The shaded area shows the range used for MAGIC
regular observations (dark and under moonlight). The dashed line shows
the linear regime. The upper axis shows the corresponding energy
threshold (after image cleaning) for observations at zenith angles
between 20$^\circ$ and 30$^\circ$ as deduced in Section~\ref{sec:energy}.}
\label{rate}
\end{center}
\end{figure}

Figure~\ref{rate} shows the dependence of the trigger rate (for a four
neighboring pixels configuration) on the DT settings, for different
illumination conditions (which produce different anode currents in the
camera). Dark observations of a galactic plane region produce anode
currents of about 1~$\mu$A and the maximum allowed current under
regular observations is 8~$\mu$A. The LONS is responsible for the
steep increase at low DT values (at $\sim$30 a.u. in the case of dark
observations). At higher DT values the rate is caused by Cherenkov
showers and a small admixture from accidental triggers caused by the
LONS and/or large amplitude afterpulses. The telescope operates at the
minimum possible DT for which the contribution of accidental triggers
is negligible. For extragalactic regions the DTs are generally set to
30 a.u., which corresponds to a pulse charge of 8-10~phe. Galactic
regions are brighter and require an increased minimum DT to the
equivalent of 11-12~phe. Even higher DT values are needed to keep the
trigger rate below the limit of the DAQ system (500 Hz) for
observations during twilight and moonlight.

\section{Observations and Data Analysis} %
\label{sec:data}

To characterize the response of the telescope under moonlight, we
observed the Crab nebula at different light conditions between January
2006 and March 2006 (see Table~\ref{datos}). The observations were
carried out in ON/OFF mode, that is, the source was observed on-axis
and observations from an empty field of view were used to estimate the
background. Two data sets, one with zenith angle between 20$^\circ$
and 30$^\circ$, and a second one between 30$^\circ$ and 40$^\circ$,
were acquired and analyzed separately. Depending on the different
moonlight levels, the resulting anode currents ranged between
1~$\mu$A and 6~$\mu$A. Correspondingly, the DT was varied between
35 and 65 units. Data from quasi-simultaneous observations during dark
time serve as reference when studying the performance of the telescope
at those moonlight level conditions.

The acquired data were processed by the standard MAGIC
analysis chain~\cite{calib,signal}. The images were cleaned using
absolute tail and boundary cuts at 10 and 5~phe, respectively. Quality
cuts based on the trigger and after-cleaning rates were applied in
order to remove bad weather runs and runs spoiled by car or satellite
flashes. The shower images were parameterized using the following
Hillas parameters~\cite{Hillas}: SIZE (total light content of the
image), WIDTH, LENGTH (second moments of the distribution of light),
DIST (distance from the image center of gravity (c.o.g) to the center
of the camera --which corresponds to the position of the observed
source), CONC (ratio between the light content of the two brightest
pixels and SIZE) and ALPHA (angle between the image major axis and the
line joining the center of the camera and the image c.o.g). Except for
the ALPHA parameter, all the other variables were combined for
$\gamma$/hadron separation by means of a Random Forest classification
algorithm~\cite{Bock,Breiman}, trained with MC simulated $\gamma$-ray
events and data from galactic areas near the source under study but
containing no $\gamma$-ray sources
\cite{mc}. The Random Forest method permits to calculate for every
event a parameter dubbed HADRONNESS, which parameterizes the purity of
hadron-initiated images in the multi-dimensional space defined by the
Hillas variables. The signal region is defined by the cuts
HADRONNESS$<$0.15 and ALPHA$<$8$^\circ$.

\section{Effect of the Moon light on the Hillas Parameters}  %

\label{sec:hillas}

\begin{figure}[!t]
\begin{center}
\epsfig{file=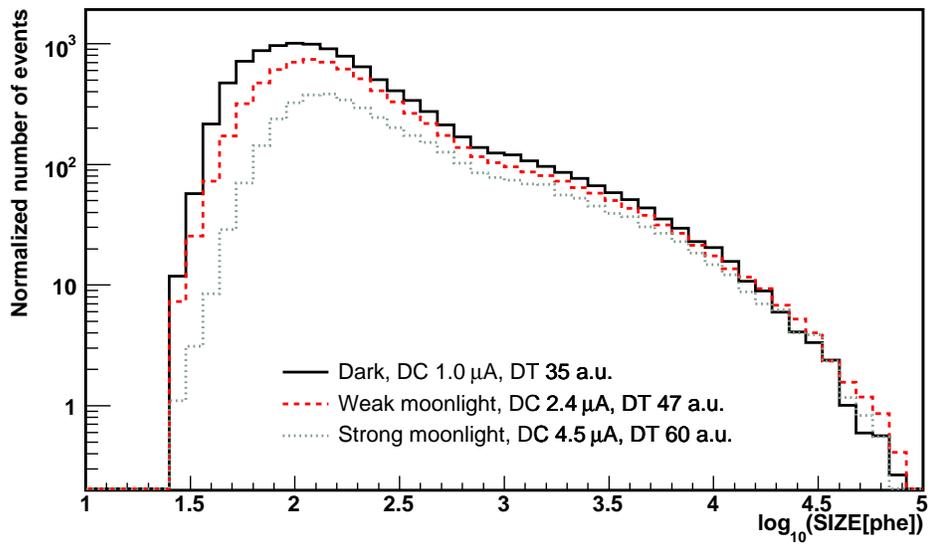,width=\textwidth}
\caption{{\it Distributions of SIZE before analysis cuts for three
Crab nebula samples acquired under different light conditions and
zenith angle between 20$^\circ$ and 30$^\circ$. The histograms have
been normalized to a common observation time. Note that the
distributions are completely dominated by hadronic events ($\sim 99\%$).}}
\label{dist_size}
\end{center}
\end{figure}

\begin{figure}[!t]
\begin{center}
\epsfig{file=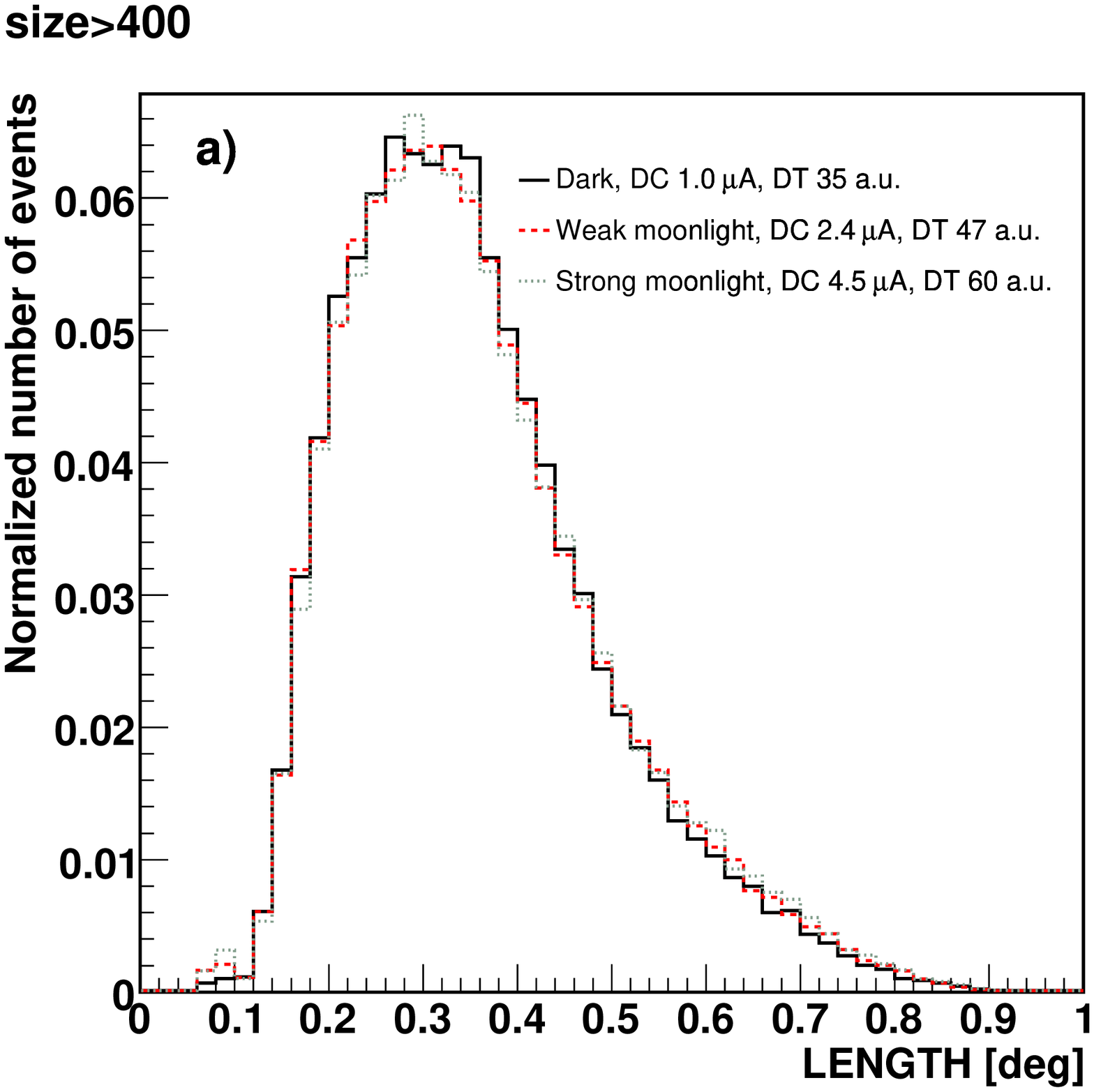,width=0.48\textwidth}
\epsfig{file=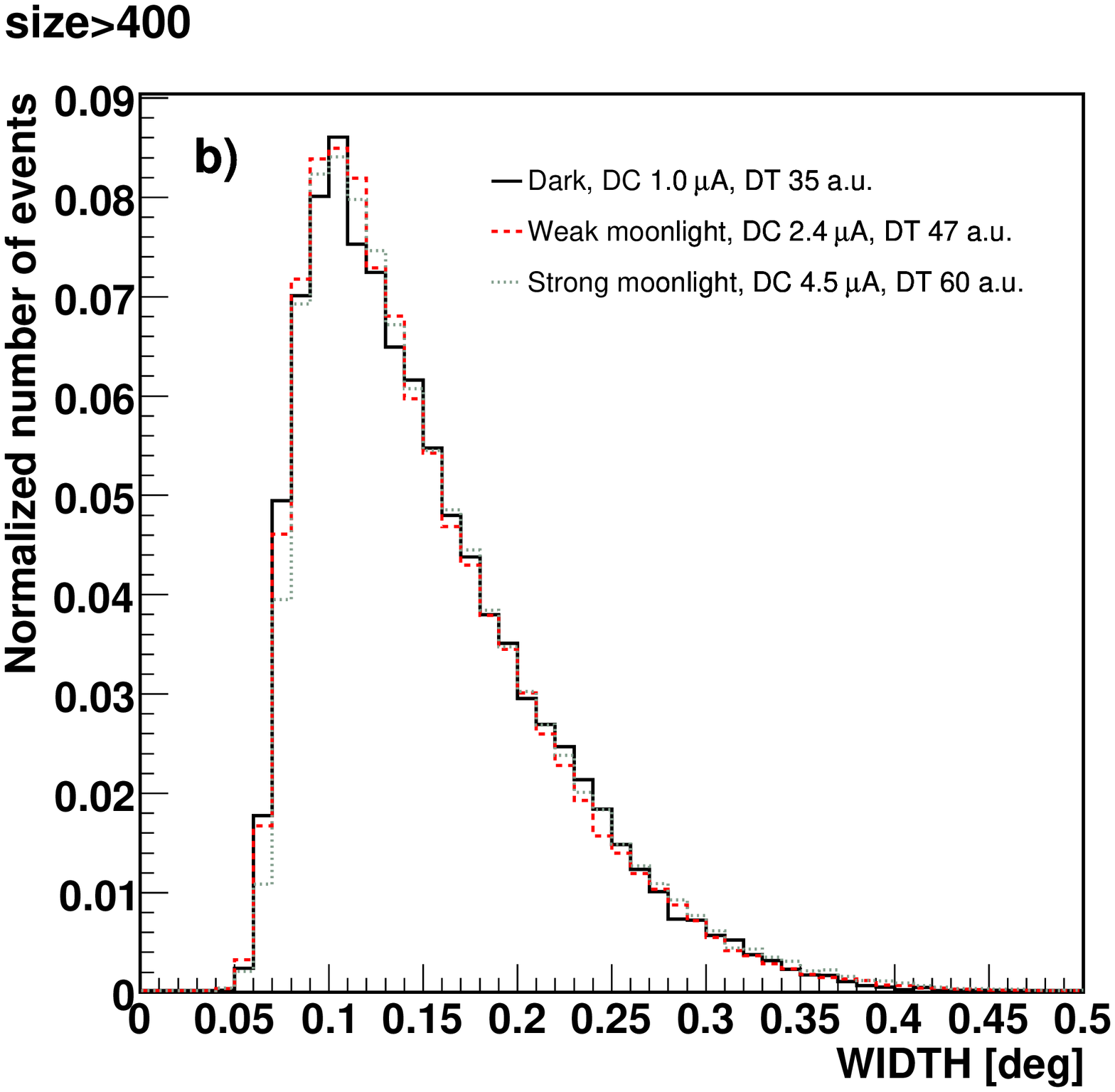,width=0.48\textwidth}
\epsfig{file=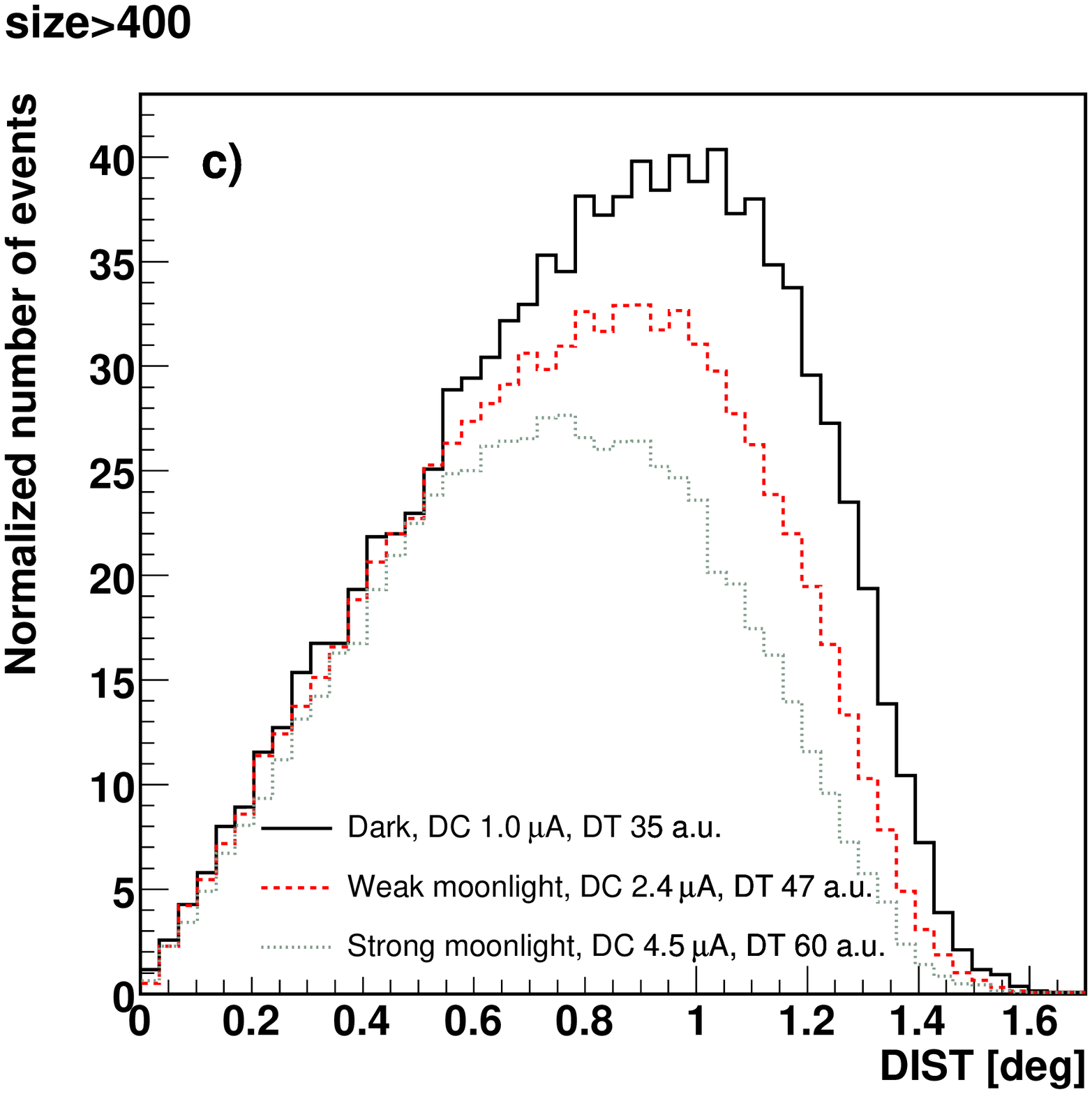,width=0.48\textwidth}
\epsfig{file=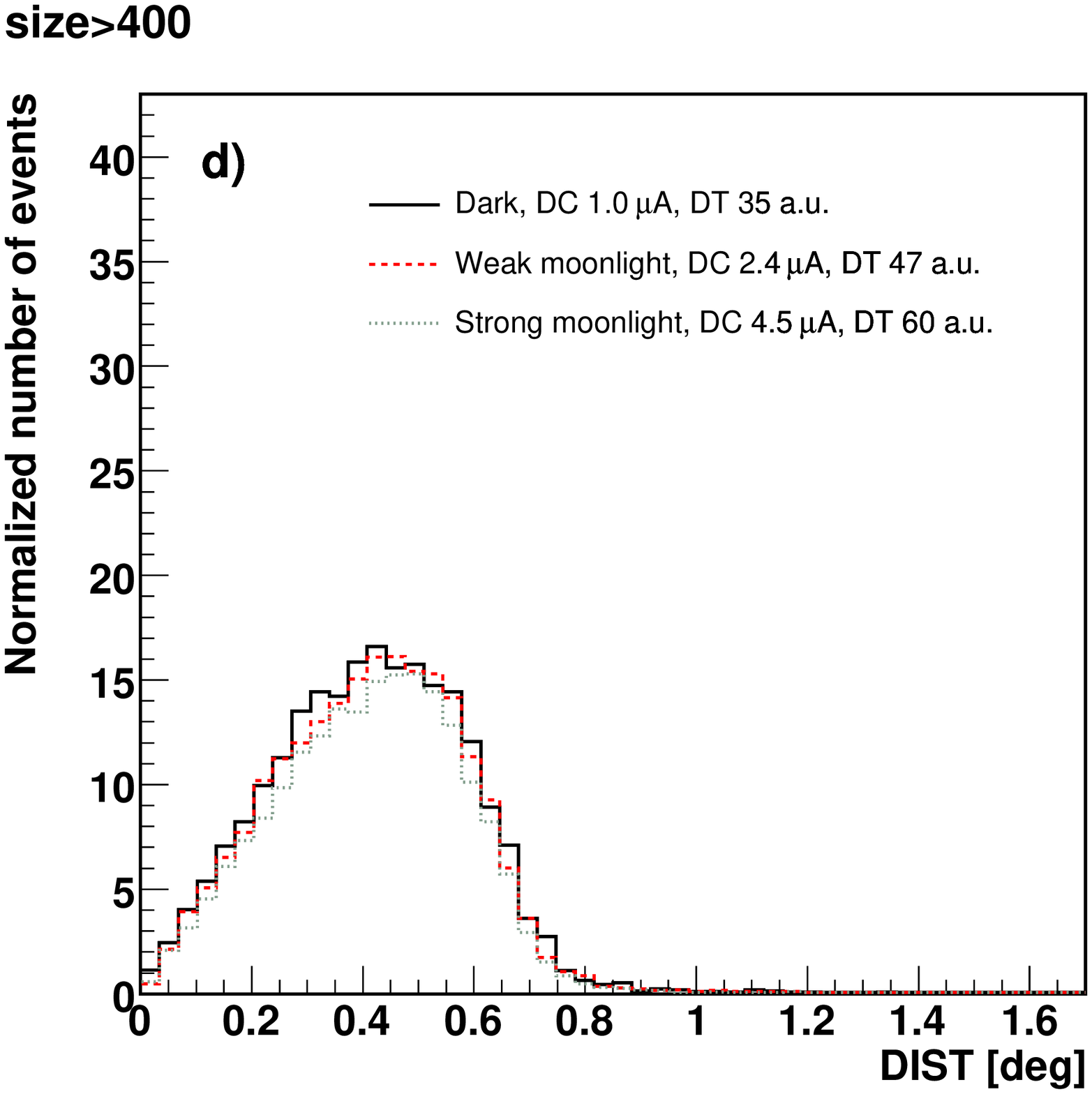,width=0.48\textwidth}
\caption{{\it Distributions of LENGTH (a), WIDTH (b) and DIST for
all recorded events (c) and for images fully contained in the inner
camera (d) for SIZE$>$400 phe. Three Crab nebula samples acquired
under different moonlight conditions and zenith angle between
20$^\circ$ and 30$^\circ$ are shown. The histograms are normalized to
a unit area in (a) and (b) and to a common observation time in (c) and
(d). Note that the distributions are completely dominated by hadronic events
($\sim 99\%$).}}
\label{dist_temp}
\end{center}
\end{figure}

Several observation samples at different Moon illuminations, hence at
different DT, were recorded for detailed studies of the impact of the
moonlight on the analysis and the telescope performance. The
distribution of the various Hillas parameters for each of these sets
are to be compared with the reference set, derived from Crab nebula
observations in dark conditions. As we increase the DT levels to
counteract accidental triggers, one depletes the SIZE distribution of
shower candidates, as expected, mostly at low values. However, we find
that a substantial number of showers with SIZE up to $10^4$~phe, i.e.\
those well above the trigger level (which is around 50 phe), are also
suppressed (see Figure~\ref{dist_size}). On the other hand, the
LENGTH, WIDTH and CONC distributions above 200~phe do not show
significant differences for events recorded either during dark nights
or in the presence of the Moon. As an example, the distributions of
LENGTH and WIDTH for SIZE$>$400~phe are shown in
Figure~\ref{dist_temp} (panels a and b). Since these parameters have
the highest discrimination power between $\gamma$ and hadron events,
we do not expect $\gamma$/hadron separation to degrade due to the
presence of the Moon. On the other hand, below 200 phe the
distributions are distorted by the different trigger threshold. A
study of the effect of the moonlight on the threshold is presented in
section~\ref{sec:energy}.

Showers with SIZE$>$1000 phe are well above the trigger level, even
for an increased DT due to the presence of the Moon. However, when the
shower impact parameter is much larger than 100~m, i.e., the Hillas
variable DIST is large, a significant fraction of the light falls
outside the trigger area (1$^\circ$ radius around the camera center
for a camera FOV of 3.5$^\circ$). In some cases, the fraction of the
shower image contained inside the trigger area will not exceed the
increased threshold for at least 4 neighboring pixels, as required for
a trigger. This effect is reflected in the DIST distribution, shown in
Figure~\ref{dist_temp}c). For a lower SIZE cut of 400~phe there is an
increased reduction of showers at large DIST ($\ge$0.8$^\circ$) values
, i.e.\ those not fully contained in the trigger area, confirming our
hypothesis. Further confirmation was obtained by two different
tests. First, the DIST distributions for events fully contained in the
trigger region were compared (Figure~\ref{dist_temp}d). In such a case
we find similar distributions, confirming our hypothesis that the
differences shown in Figure~\ref{dist_temp}c) are coming from events
whose image is only partially contained in the trigger region. A last
test was performed by observing Crab in dark conditions
(DC$\sim$1.1~$\mu$A), but with increased DTs. In this case we found
similar inefficiencies as those shown in Figure
\ref{dist_temp}c). Therefore we can conclude that the change of the DIST
distribution is not related to the mean DC current (i.e. with the
camera illumination) but only to the DT level.

These results show that moonlight does not distort the images from
Cherenkov showers --for an image cleaning based on the pixels'
absolute light content, as that used in MAGIC. This has two important
consequences. At first, the analysis based on the Hillas parameters
does not have to be adapted for data acquired under moonlight, and in
particular the $\gamma$/hadron separation power is not reduced for
these kind of observations. Secondly, the differences that we find in
the event rates and the DIST distributions are exclusively due to the
fact that the DTs were increased to keep a low rate of accidental
events, together with the fact that the trigger area does not span the
whole camera.  It is important to remark that this is a merely
technical issue, imposed by the intrinsic maximum rate the DAQ system
can handle (about $\sim 500$~Hz in the present MAGIC
configuration). For this study we followed the conservative approach
of increasing the DTs, and hence the dependence of the telescope
response on the level of Moon illumination will be parameterized as a
function of DT. However, with a fast enough DAQ system, we could keep
constant the DTs and deal with the increased amount of accidental
events produced by the moonlight during the off-line analysis, or by a
more developed second-level trigger system.

\section{Telescope Performance} %
\label{sec:results}

As was shown above, shower images are not distorted by the
moonlight. However, the performance of the telescope is modified for
this kind of observations with respect to dark conditions, due to the
increase of the DTs. As is shown in Figure~\ref{dist_size}, there is a
reduction in the collection area over a wide range of energies. In
principle, this effect could be taken into account by proper MC
simulation of the different trigger conditions. In practice, however,
the DTs are not fixed during the MAGIC observations, but change
dynamically to compensate for short term variations of the camera
illumination caused by the movement of the Moon and the source along
the sky, as well as for variations of the position of stars in the
MAGIC camera. This makes such a simulation a difficult task. Instead,
we use Crab nebula observations to estimate the efficiency of detecting
$\gamma$-rays, for every DT and SIZE, relative to the values for dark
observations. These values are used during the off-line analysis
together with the MC simulation with standard DTs to calculate the
correct collection areas for every DT and SIZE. We also compute the
effect of the moonlight on the telescope sensitivity (the minimal flux
detectable with $5\sigma$ significance in 50~hours of observations),
and on the energy threshold. Note that we will in the following use
the level of DT as an equivalent measure of the moonlight.

\subsection{$\gamma$-ray Detection Efficiency and Telescope Sensitivity} 

\begin{figure}[!t]
\begin{center}
\epsfig{file=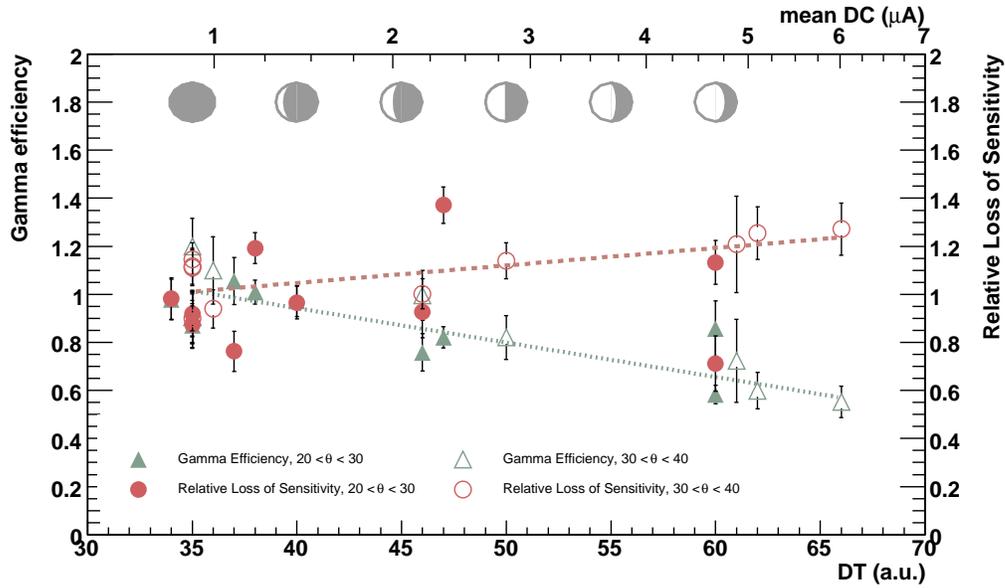,width=\textwidth}
\caption{{\it Relative $\gamma$-ray detection efficiency (green triangles, left
axis) and sensitivity (red circles, right axis) as a function of DT
(SIZE$>$ 400~phe), for zenith angle bins $[20^\circ,30^\circ]$ (filled
markers) and $[30^\circ,40^\circ]$ (empty markers) measured from Crab
nebula observations. The sketches showing the Moon phase are
meant to guide the reader, since the camera illumination does not only
depend on the phase, but also on factors such as the angular distance
between source and Moon, etc.}}
\label{resultsdata}
\end{center}
\end{figure}

Observations of the Crab nebula are divided into different samples
according to the observation date and the DT value. For each of the
samples we get a measurement of the $\gamma$-ray rate ($R$), i.e. the
number of excess events per unit time. We find that, for a given SIZE
range, the dependence of $R$ with the DT is well described by a linear
function:
\begin{equation} 
R = R_0 \left(1 - S_\epsilon \left(DT - DT_0\right)\right) 
\label{eq:efficiency}
\end{equation} 
where $R_0$ is a normalization factor, $S_{\epsilon}$ is the {\it
efficiency loss rate}, and $DT_0$ is a reference DT value that, for
convenience, is chosen as the one used in dark observations, i.e.
$DT_0=35$. We present the results for $\epsilon \equiv R/R_0$,
i.e. the $\gamma$-ray detection efficiency with respect to the case
$DT=DT_0=35$ (dark observations). The results for SIZE$>$400 phe and
the two considered zenith angle samples ($[20^\circ,30^\circ]$ and
$[30^\circ,40^\circ]$) are shown in Figure~\ref{resultsdata}. The fit
parameters obtained for both zenith angles are compatible within
statistical errors. This allows us to perform a combined fit for both
samples. We obtain the following expression for the $\gamma$-ray
detection efficiency:
\begin{equation}
\epsilon = 1 - (1.41\pm 0.32)\times10^{-2}(DT-DT_0)
\label{eq:effloss}
\end{equation}
The error quoted is obtained from the fit and it does not include any
correlation with the normalization factor. The overall uncertainty
obtained from the residuals of the individual data points is 0.1. This
most likely comes from variations in the experimental conditions
(weather, hardware, etc) in the different observation nights, and can
be regarded as a measure of the systematic point-to-point uncertainty
when measuring light-curves within the total observation time of our
analysis, that is, three months. The results show that the
$\gamma$-ray detection efficiency for events above SIZE $>400$~phe is
reduced with the moonlight brightness with a rate of 1.41$\%$ per DT
count, implying that, for high illumination of the camera, detection
efficiency losses up to 50$\%$ are expected.  Moreover, this result
seems to be independent of the studied zenith angle range. A study of
the different dependence of the efficiency as a function of SIZE will
be discussed in Subsection~\ref{sec:energy}.

We have also measured the loss of sensitivity ($s$) produced by the
increase of the DTs. In dark conditions the flux sensitivity ($s_0$)
of the MAGIC telescope above 400 phe is 2.5$\%$ of the Crab nebula
flux~\cite{sensi}. The relative sensitivity ($s/s_0$) is computed
using Crab nebula observations under different moonlight conditions by:
\begin{equation}
s/s_0 = \frac{N_\gamma^0 \sqrt{t N_\textrm{bkg}}}{N_\gamma
\sqrt{t^0 N_\textrm{bkg}^0}}
\end{equation}
where $N_\gamma$ and $N_\textrm{bkg}$ are, respectively, the number of
excess and background events after analysis cuts (see
Section~\ref{sec:data}) for an observation lasting a time $t$. The
upper index 0 stands for the values for the DT=35. The dependence of
the sensitivity loss as a function of DT for a SIZE cut $>$400~phe is
shown in Figure~\ref{resultsdata}. It is well fitted by a linear
function, parameterized as $s/s_0 = 1 + S_s (DT-DT_0)$, where
$S_s$ is referred to as {\it sensitivity loss rate}, that is:
\begin{equation}
s/s_0 = 1 + (6.3\pm 1.6)\times10^{-3}(DT-DT_0)
\label{eq:sensi}
\end{equation}

\begin{figure}[!t]
\hspace{0.5cm}
\begin{center}
  \epsfig{file=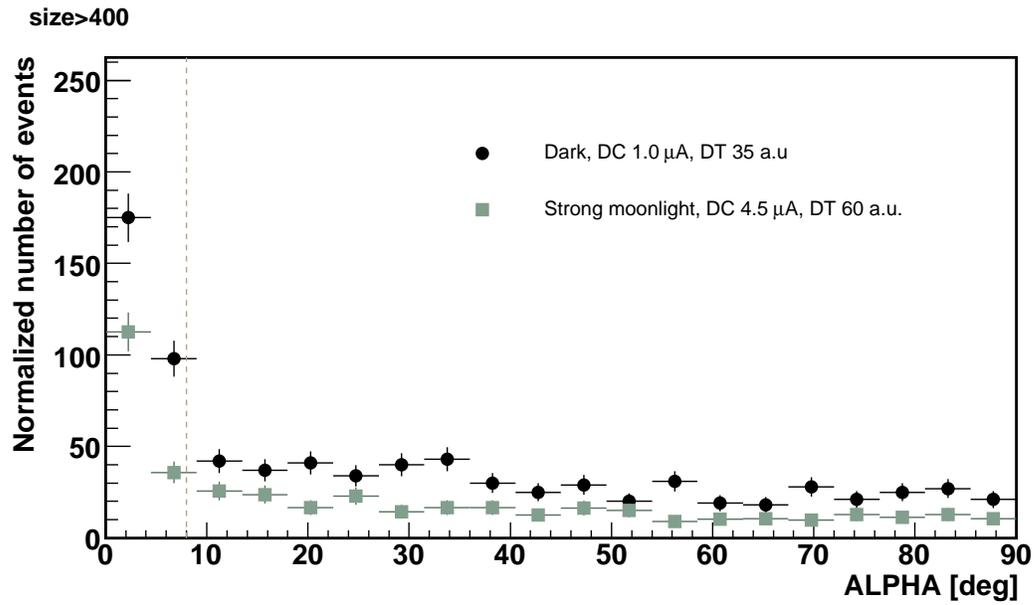,width=\textwidth}
\caption{{\it ALPHA distributions (SIZE$>$400~phe, zenith angle
between 20$^\circ$ and 30$^\circ$) for dark observations and
observations under strong moonlight. The distributions are
normalized to a common observation time. The vertical dashed line shows the
signal selection cut applied in our analysis.}}
\label{fig:alpha}
\end{center}
\end{figure}

Therefore, a loss of sensitivity of $(6.3 \pm 1.6)\permil$ per DT unit
is observed for a cut SIZE$>$400~phe. This value can be compared with
the $(1.4 \pm 0.32)\%$ loss in the $\gamma$-ray detection efficiency obtained in
Equation~\ref{eq:effloss}. For instance, for a rather high moonlight
brightness, e.g. for DT=60, we get a $\gamma$-ray detection efficiency of 65$\%$
but the sensitivity decreases only by 15$\%$ (e.g.\ from a 2.5$\%$ to a
2.9$\%$ of the Crab nebula flux). This is due to the fact that signal
and background rates both are equally reduced by an increase of the
trigger threshold. The effect is further illustrated in
Figure~\ref{fig:alpha}, where the distributions of ALPHA for dark and
strong moonlight observations are compared (in this case the reduction
of signal and background rates is a factor of $\sim 2$). One expects
the loss of sensitivity to go as the square root of the loss of
$\gamma$-ray rate, which is, within statistical errors, what we find
in our data.

\begin{figure}[!t]
\begin{center}
  \epsfig{file=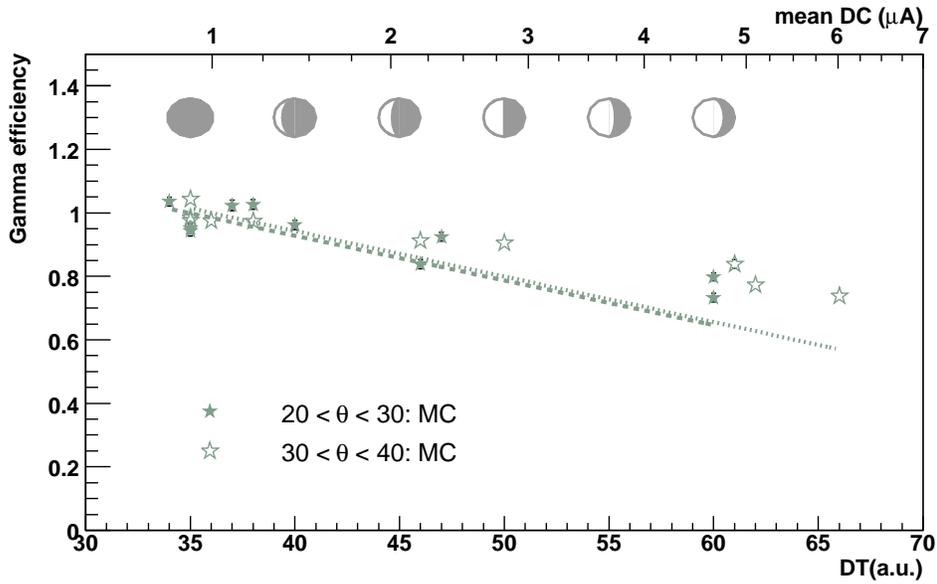,width=\textwidth}
\caption{{\it Relative $\gamma$-ray detection efficiency as a function of DT for
SIZE$>$ 400~phe as obtained from Crab nebula data (lines) and from MC
simulations (stars) for zenith angle samples $[20^\circ,30^\circ]$
(empty markers) and $[30^\circ,40^\circ]$ (filled markers). The
sketches showing the Moon phase are meant to guide the reader,
since the camera illumination does not only depend on the phase, but
also on factors such as the angular distance between source and Moon,
etc. }}
\label{mc}
\end{center}
\end{figure}

The results described above are compared with Monte Carlo (MC)
simulations, for which the background light fluctuations and pixel
trigger thresholds are increased in agreement with the values observed
in the data. In Figure~\ref{mc} we show the results of the fit to the
$\gamma$-ray detection efficiency as determined from the different Crab nebula
observations (from Figure~\ref{resultsdata}), together with the
results of the MC simulation. The result shows agreement in the general
trend. The evident discrepancy between data and MC at large DT can be
attributed to the fact that the simulation of the MAGIC trigger is not
yet optimized for this type of observations. Note
that the results shown in this paper rely on real data events only, and
therefore are not affected by this discrepancy.

\subsection{Dependence on SIZE and Energy} 
\label{sec:energy}

\begin{figure}[!t]
\begin{center}
  \epsfig{file=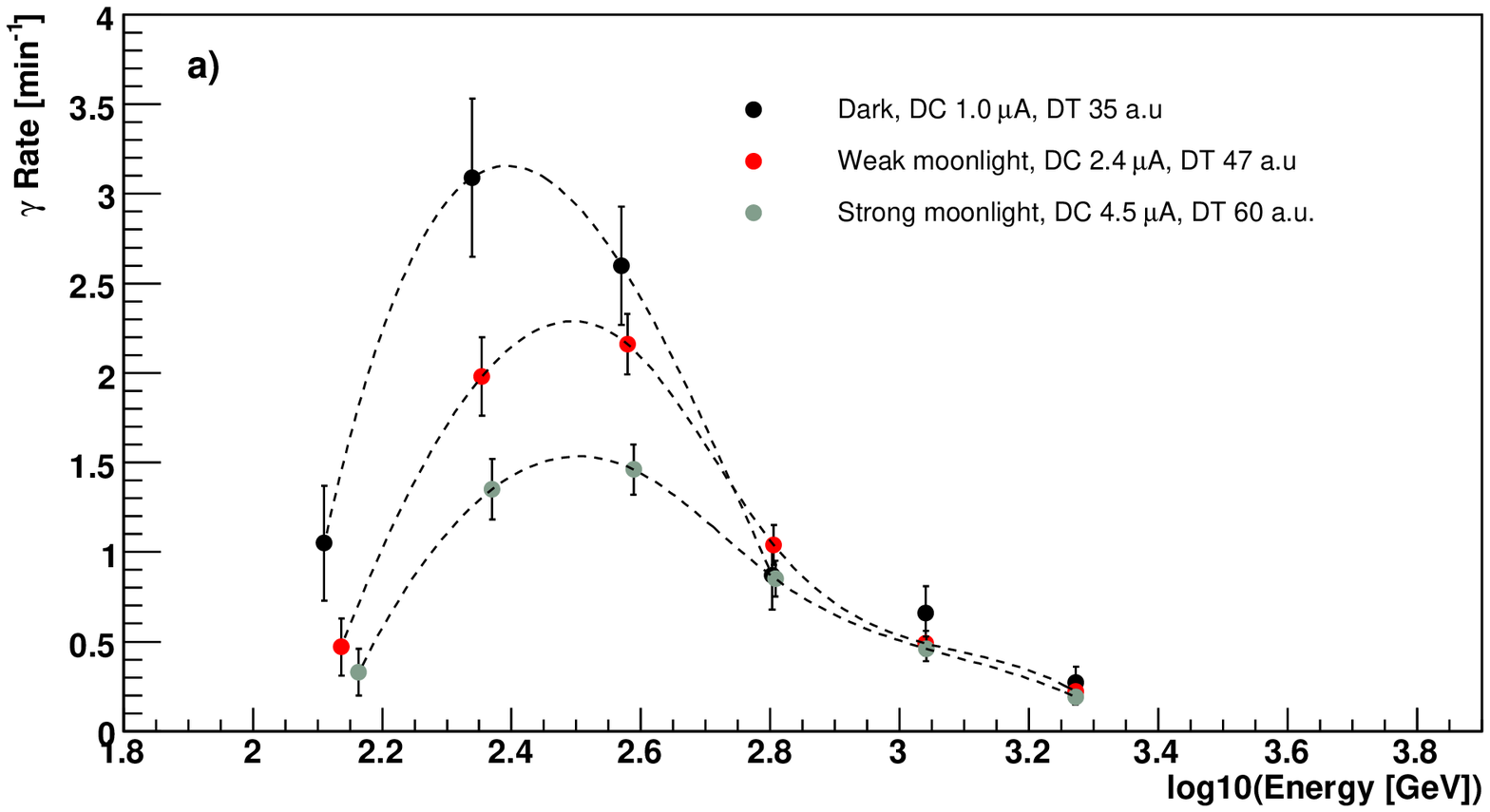,width=\textwidth}
  \epsfig{file=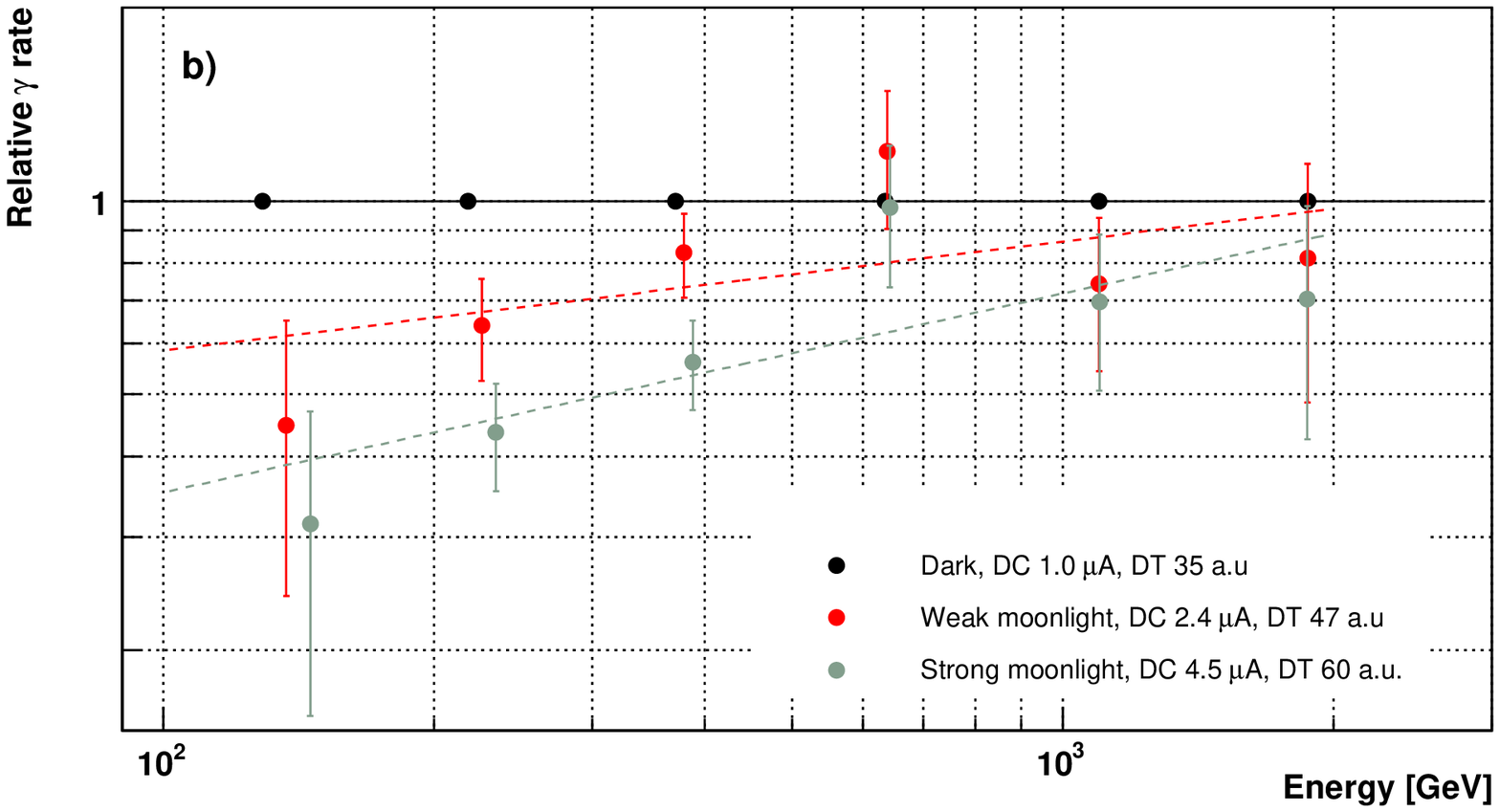,width=\textwidth}
\caption{\it Absolute (a) and relative (b) $\gamma$-ray rate after
analysis cuts (HADRONNESS$<$0.15, ALPHA$<$8$^\circ$) as a function of
the estimated energy. Results from 
observations of the Crab nebula under three different moonlight
intensities and zenith angle between 20$^\circ$ and 30$^\circ$ are
shown. In (a) the dashed lines are a polynomial interpolation of the data
points and are meant to guide the eye only. In (b) the data points of
dark observations serve as a reference; the dashed lines show the
result of fitting a power law to the different data sets.}
\label{fig:raw_spec}
\end{center}
\end{figure}

The results obtained in the previous section are valid for the
integral flux above SIZE$>$400~phe. However, the loss of efficiency
and sensitivity are expected to depend strongly on SIZE and hence on
the energy. Figure~\ref{fig:raw_spec} shows the $\gamma$-ray rate as
a function of the energy (both in absolute value and relative to the
values for dark observations), for three different moonlight
intensities. As expected, the loss is larger for more intense camera
illuminations mainly at low energy values, whereas for high energies
we find no significant differences. In order to quantify the impact of the
moonlight in the physical parameters extracted from the MAGIC
observations (that is, the flux normalization and spectral shape), 
we fit a power-law function ($R = (1-R_{400}) (\frac{E}{400
\textrm{GeV}})^{-\Delta \alpha}$) to the relative $\gamma$-ray rate
(Figure~\ref{fig:raw_spec}b) for the different moonlight intensities.
The normalization is chosen at 400~GeV to minimize the correlation
between the errors of the two free parameters, namely: $R_{400}$
and $\Delta \alpha$. The fit yields $R_{400}=0.26\pm 0.07$, $\Delta
\alpha = -0.17\pm 0.12$ for a weak camera illumination and $R_{400}=
0.46\pm 0.05$, $\Delta \alpha = -0.31\pm 0.12$ for a strong camera
illumination.  $R_{400}$ and $\Delta \alpha$ can be regarded as the
systematic effect that would be introduced in the determination of
the spectral parameters if not treating properly the moonlight-related changes.
They can be compared to the systematic
errors coming from other sources: $\sim 10\%$ for the flux normalization
(see Section~\ref{sec:results}) and 0.2 for the spectral index, to
stress the importance of this study to measure the spectral
parameters.

\begin{figure}[!t]
\begin{center}
\epsfig{file=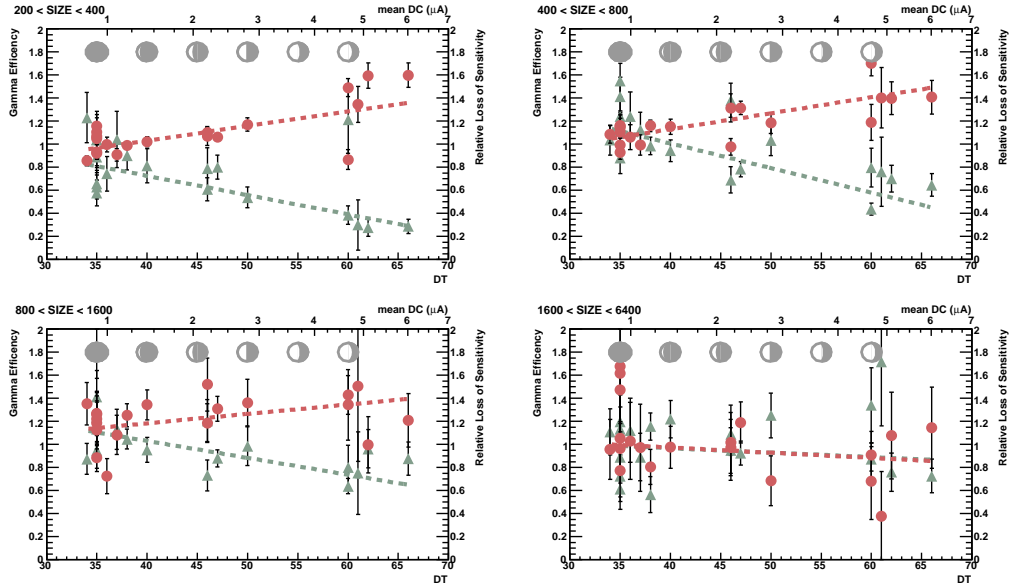,width=\textwidth}
\caption{{\it Effect of the moonlight on the $\gamma$-ray detection efficiency
(green) and sensitivity (red) as a function of DT for different
SIZE bins, measured from Crab nebula observations at zenith
angles between 20$^\circ$ and 30$^\circ$. The best fits to a
linear function are also shown. The sketches showing the Moon phase
are  meant to guide the reader, since the camera illumination does
not only depend on the phase, but also on factors such as the angular
distance between source and Moon, etc.}}
\label{size_lowzd}
\end{center}
\end{figure}

In order to understand the dependence of the $\gamma$-ray detection
efficiency on the energy we have the same study for four bins of SIZE,
namely [200,400], [400,800], [800,1600] and [1600,6400]~phe, which
roughly correspond to the energy ranges [150,300], [300,600],
[600-1000] and $>$~1000~GeV, respectively, for low zenith angle. The
$\gamma$-ray efficiencies as a function of the moonlight brightness
(using the DT setting as the equivalent measure) are shown in
Figure~\ref{size_lowzd} for zenith angles between 20$^\circ$ and
30$^\circ$. The data sample with zenith angles between 30$^\circ$ and
40$^\circ$ was also analyzed with similar
results. Equation~\ref{eq:efficiency} describes reasonably well the
data in all SIZE bins considered. The efficiency loss rate
($S_\epsilon$) decreases as SIZE increases since the probability of
showers passing a high trigger level is higher for larger images. For
high enough values of SIZE the $\gamma$-ray detection efficiency
remains constant with DT (at least up to DT=60).

\begin{figure}[t]
\begin{center}
\epsfig{file=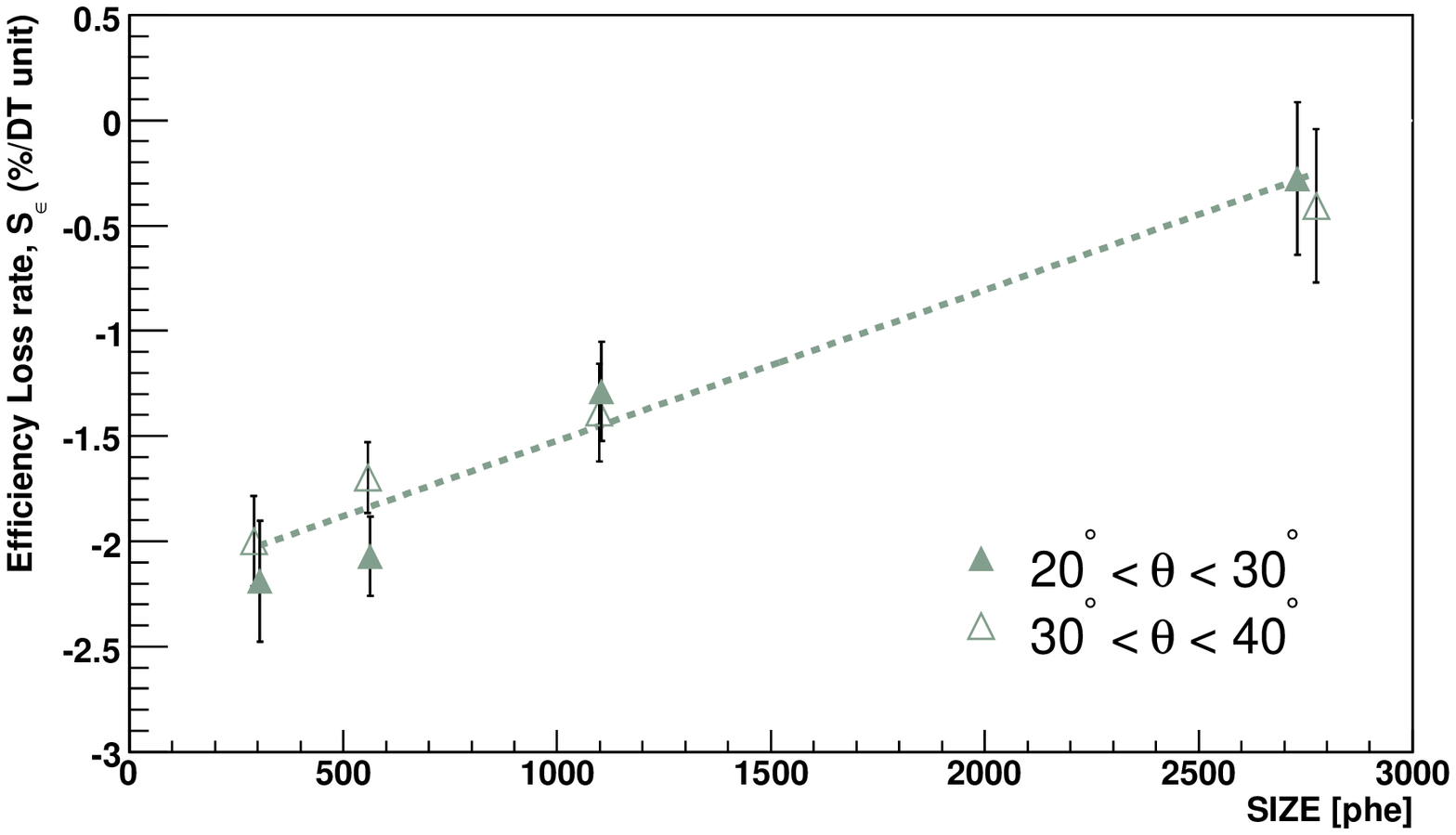,width=\textwidth}
\caption{{\it $\gamma$-ray detection efficiency loss rate ($S_\epsilon$ in
$\%$/DT unit) as a function of SIZE obtained from Crab nebula
observations at zenith angles between 20$^\circ$ and 40$^\circ$. The
line represents the result of a linear fit to the data points. Note
that the linear dependence must level off above
SIZE$\approx$3000~phe.}}
\label{SlopeAll}
\end{center}
\end{figure}

We observe no significant differences in the $\gamma$-ray detection
efficiency between medium and low zenith angles, that is, the effect of
the Moon does not depend on the zenith angle of the considered
source. Figure~\ref{SlopeAll} shows $S_\epsilon$ as a function of
SIZE. The x-values of the shown data are the peak of the SIZE
distribution for the different considered bins, obtained from MC
simulated $\gamma$-ray events based on a Crab-like power-law spectrum
and nominal DT. Up to SIZE=3000~phe we find a linear dependence that
can be parameterized by:

\begin{equation} 
S_\epsilon  =  (2.24\pm0.13)\times10^{-2} - (7.2\pm 1.2)\times10^{-6}  \textrm{ SIZE [phe]}
\label{eq:efi_size}
\end{equation}

\begin{figure}[!t]
\begin{center}
  \epsfig{file=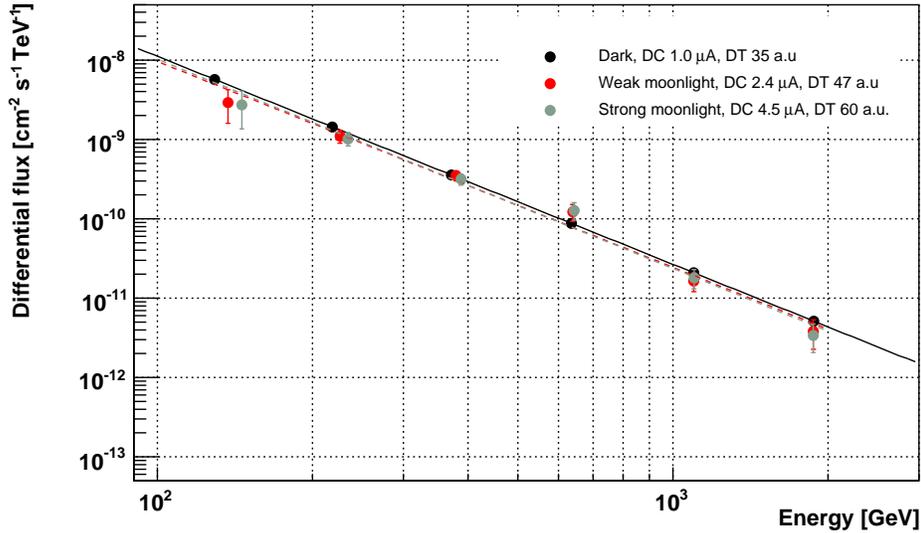,width=\textwidth}
\caption{\it Differential energy spectrum for the Crab
nebula observed under three different moonlight intensities and zenith
angle between 20$^\circ$ and 30$^\circ$. The dashed lines show the
result of fitting a power law to the data points.}
\label{fig:corr_spec}
\end{center}
\end{figure}

The quoted errors are obtained from the fit, and include the
correlation between the two free parameters. This is a very useful
result since, together with Equation~\ref{eq:efficiency}, it allows
one to compute the $\gamma$-ray detection efficiency loss for any SIZE and DT
(i.e.\ for all tested moonlight conditions), and hence to correct the
$\gamma$-ray fluxes during the off-line analysis, with no need of
generating different MC samples for the different moonlight
conditions. The results are rather independent of the zenith angle, at
least for values below 40$^\circ$. Figure~\ref{fig:corr_spec} shows
the measured energy spectra for the Crab nebula from observations
under weak and strong moonlight after applying the SIZE-DT dependent
correction factors computed from Equations~\ref{eq:efficiency} and
\ref{eq:efi_size}. The fit of a power-law ($F = F_{400} (\frac{E}{400
\textrm{GeV}})^{-\alpha}$ 10$^{-10}$ cm$^{-1}$ s$^{-1}$ TeV$^{-1}$)
yields $F_{400}=2.63\pm 0.24$, $\alpha = 2.59\pm 0.07$ for weak camera
illumination, and $F_{400} = 2.65\pm 0.26$, $\alpha = 2.63 \pm 0.08$
for strong camera illumination. These values are statistically
compatible with those obtained for dark observations ($F_{400}=2.9
\pm 0.3$, $\alpha = 2.58\pm 0.16$)~\cite{sensi}.

\begin{figure}[t]
\begin{center}
\epsfig{file=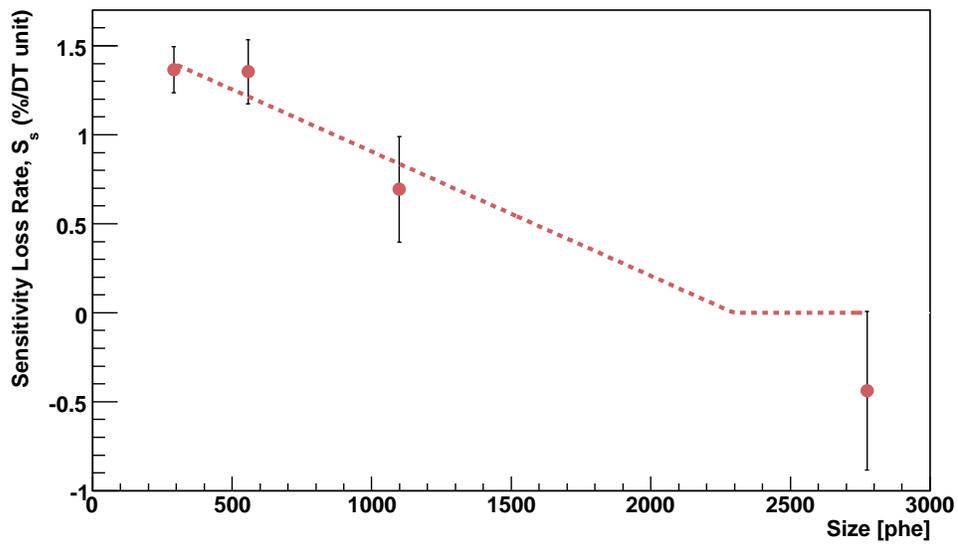,width=\textwidth}
\caption{\it Sensitivity loss rate ($S_s$ in $\%$/DT unit) as a
function of SIZE obtained from Crab nebula observations at zenith
angle between 20$^{\circ}$ and 40$^{\circ}$. The line represents the
result of a linear fit to the data points. Note that the linear
dependence must level off above SIZE$\approx$2500~phe.}
\label{SlopeAllSensi}
\end{center}
\end{figure}

A similar analysis has been carried out to evaluate the differential
sensitivity loss as a function of energy, i.e.\ as a function of
SIZE. Figure \ref{size_lowzd} shows the sensitivity loss rate as a
function of DT for the four aforementioned SIZE bins. For lower SIZE
the sensitivity degrades with increasing DT reaching a maximum
relative value of 50$\%$, while for larger SIZE it remains roughly
constant with increasing DT. The sensitivity loss rate as a function
of SIZE is shown in Figure~\ref{SlopeAllSensi}. A linear fit to the
data yields:
\begin{equation}
S_s  =  (1.63\pm0.14)\times10^{-2} - (7.4\pm 1.8)\times10^{-6}  \textrm{ SIZE [phe]}
\label{eq:sensi_size}
\end{equation}

We tested the dependences quoted in Equations~\ref{eq:efi_size} and
\ref{eq:sensi_size} under different HADRONNESS and ALPHA cuts,
obtaining in all cases similar results in the efficiency and
sensitivity loss rate within the statistical and systematic
uncertainties.

\begin{figure}[!t]
\begin{center}
\epsfig{file=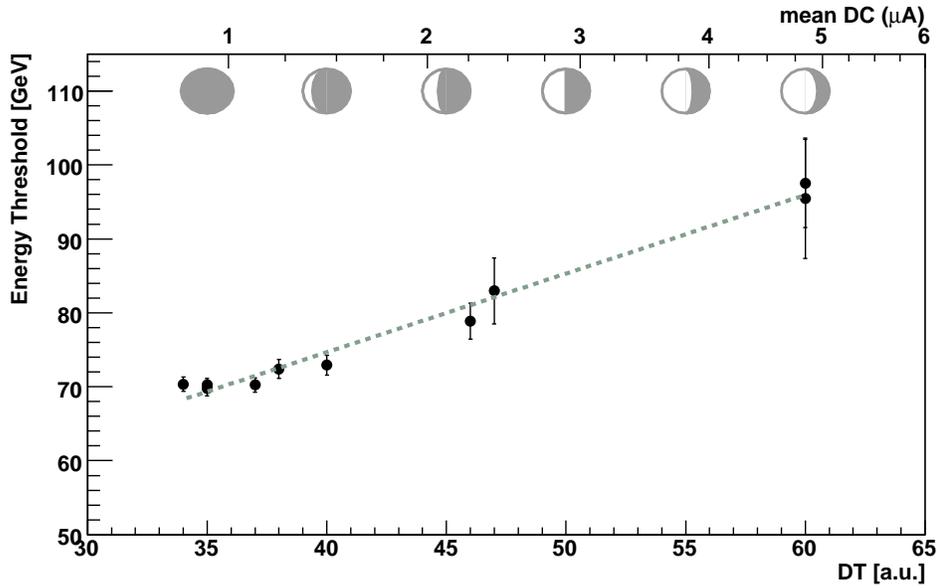,width=\textwidth}
\caption{{\it Energy threshold after image cleaning
as a function of DT obtained from MC simulated $\gamma$-ray events
(for zenith angle between 20$^\circ$ and 30$^\circ$). The top axis
shows the typical mean DC for a chosen DT value. The sketches showing
the Moon phase are meant to guide the reader, since the camera
illumination does not only depend on the phase, but also on factors
such as the angular distance between source and Moon, etc.}}
\label{threshold}
\end{center}
\end{figure}

Finally, it is important to understand the influence of the moonlight
on the energy threshold. We define the energy threshold as the peak of
the energy distribution of all events after image cleaning and before
analysis cuts. The energy is estimated assuming the correspondence
SIZE/energy obtained from a MC simulated $\gamma$-ray sample (zenith
angle $20-30^\circ$). The dependence of the energy threshold is well
described by the following linear function (see
Figure~\ref{threshold}):
\begin{equation}
E_\textrm{th}=(69.3\pm 0.4)+(1.06\pm 0.03)~(DT-DT_0) ~~~~ [\textrm{GeV}]
\end{equation}
where the errors quoted are obtained from the fit and include the
correlation between the two free parameters. From the above expression,
the energy threshold increases $\sim 1$~GeV per DT unit. An increase
of the threshold from $\sim 70$ to $\sim 95$~GeV at a mean zenith
angle of 25$^\circ$ is observed at the maximally allowed camera
illumination. This increase is relatively marginal, and it has to be
noted again that it is due to the increase of the DTs, and hence only
indirectly to the increase in the camera illumination.

\section{Conclusions}  %

\label{sec:concl}

The camera of the MAGIC telescope was designed to run under moderate
moonlight or twilight by means of a reduced gain of the PMT dynode
amplification system. MAGIC PMTs have been shown to operate under
illumination up to thirty-five times stronger than the dark sky with
no substantial degradation of their performance. The telescope has
routinely operated in moonlight and twilight conditions for the past
two years.

We have characterized MAGIC's response in observations under moderate
moonlight or twilight, using Crab nebula observations. The DTs were
increased to keep the rate of accidental triggers produced by the
moonlight roughly constant. We have found that, under such
experimental conditions, we can characterize the performance of the
telescope in terms of the selected DT value. In comparison, the noise
induced by the moonlight itself has been shown to contribute in a
negligible way. The distributions of the Hillas variables --and hence
the $\gamma$-hadron separation power-- are in good agreement. This
makes the data taken under moonlight very simple to analyze, since no
special treatment is required at the calibration, image cleaning and
$\gamma$/hadron separation stages. Another important consequence is
that, with a DAQ system capable of handling a much higher rate, no
changes on the DTs are required, and the images produced by Cherenkov
showers would still be basically of the same quality compared to those
recorded during dark nights. During MAGIC operation, the DTs are set
to a minimum value ensuring both, a low rate of accidental trigger
events, and a minimal decrease of the collection area. Our results
suggest that this is a somewhat conservative approach, and that the
only real limitation is given by the intrinsic maximum rate of the DAQ
system. The possibility of changing the second level of trigger (e.g.\
to a five neighboring pixels configuration) to reject the accidental
events without increasing the DTs is left for a future study.

The increase in the DT produces a marginal increase of the telescope's
energy threshold, even for the most intense illuminations tested in
this study. On the other hand, we have found important effects in the
collection area in a wide range of energies, attributable to the
increase of the DTs and a relatively small trigger area, that have
been parameterized in terms of losses in the $\gamma$-ray detection
efficiency. The sensitivity is comparatively less affected, due to the
fact that hadrons also suffer from the same detection efficiency
losses. The loss of sensitivity that we measure is compatible with
zero at high energies, while for low energy/SIZE values the effect
becomes important. The $\gamma$-ray detection efficiency decreases
linearly with increasing DT levels and decreasing SIZE. Such
dependences have been fully characterized using observations of the
Crab nebula. When analyzing MAGIC's moonlight or twilight data, the
only concern is to correct the number of $\gamma$-ray events in a
spectrum or light curve by the efficiency corresponding to the given
DT and SIZE values, with no need of special MC simulations for the
large (in principle, infinite) variety of moonlight conditions. It is
also shown that these results are valid independent of particular
analysis cuts, provided they are based on the Hillas image
parameterization. The results are also independent of the zenith angle
of the observation, at least up to 40$^\circ$.

MAGIC now performs regular observations under moderate moonlight
during cycle II (May 2006-April 2007), so far with a cumulated time of
$\sim200$~h taken so far (Feb.\ 2007), representing $\sim 25\%$ of the
total observation time during the same period. Some of these
observations include the first unidentified TeV source
TeV~J2032+40~\cite{SNRs}, SNRs such as Cassiopeia~A, and variable
sources such as the binary system LSI~+61~303~\cite{LSI} and AGNs like
Mkn~501 and Mkn~421. Apart from the obvious gain in terms of duty
cycle, the possibility to extend the observations also in the presence
of the Moon has an important relevance in the study of variable
sources. The increase of the observation window also allows a better
overlap with other astronomical instruments when participating in
multi-wavelength campaigns, and hence in the understanding of the
physical processes governing the entire electromagnetic emission of
variable sources. In particular, observations under moonlight have
been of crucial importance in the study of the near AGN Mkn 501 with
MAGIC, that has unveiled variability over exceptionally short time
scales~\cite{Mkn501}, or in the determination by MAGIC of the periodic
nature of LSI +61 303 at VHE. Last but not least, an enlargement of
the observation window increases the probability of detecting Gamma
Ray Bursts~\cite{GRB} at VHE by MAGIC.

\section{Acknowledgments}
We would like to thank the IAC for the excellent working conditions at
the Observatorio del Roque de los Muchachos in La Palma.  The support
of the German BMBF and MPG, the Italian INFN and the Spanish CICYT is
gratefully acknowledged. This work was also supported by ETH Research
Grant TH~34/04~3 and the Polish MNiI Grant 1P03D01028.


\newpage
  
\begin{table}[!t] 
\caption{{\it Crab nebula observations (January 2006 to March 2006) }}
\label{datos}
\begin{center}
\begin{tabular}{lll} 
\hline
Zenith angle & 20$^\circ<\theta<$30$^\circ$   &  30$^\circ<\theta<$40$^\circ$  \\
\hline 
Observation time (min) & 488  & 297 \\
Mean DC range ($\mu$A)  & 1.0 - 4.5 & 1.0 - 5.2 \\ 
Mean DT range  & 35 - 60 & 35 - 66  \\
\hline
\end{tabular}
\end{center}
\end{table}

\end{document}